\documentclass[aps,prb,twocolumn,groupedaddress,showpacs]{revtex4-1}
\usepackage{amsmath,amsthm,amssymb}
\usepackage[dvips]{graphicx}

\begin{document}

\newcommand{\bi}[1]{\ensuremath{\boldsymbol{#1}}} 

\title{Excited configurations of hydrogen in the BaTiO$_{3-x}$H$_x$
perovskite lattice\\ associated with hydrogen exchange and transport}

\author{T.~U.~Ito$^{1}$} 
\author{A.~Koda$^{2}$}
\author{K.~Shimomura$^{2}$} 
\author{W.~Higemoto$^{1,3}$} 
\author{T.~Matsuzaki$^{4}$}
\author{Y.~Kobayashi$^5$}
\author{H.~Kageyama$^{5}$}
\affiliation{$^1$Advanced Science Research Center, Japan Atomic Energy
 Agency, Tokai, Ibaraki 319-1195, Japan}
\affiliation{$^2$Institute of Materials Structure Science, High Energy
Accelerator Research Organization (KEK), Tsukuba,
Ibaraki 305-0801, Japan}
\affiliation{$^3$Department of Physics, Tokyo Institute of Technology, 
Meguro, Tokyo 152-8551, Japan}
\affiliation{$^4$Advanced Meson Science Laboratory, RIKEN, Wako
351-0198, Japan}
\affiliation{$^5$Department of Energy and Hydrocarbon Chemistry,
Graduate School of Engineering, Kyoto University, Nishikyo-ku, Kyoto
615-8510, Japan}

%\author{T.~U.~Ito$^{1,2}$} 
%\author{A.~Koda$^{2,3}$}
%\author{K.~Shimomura$^{2,3}$} 
%\author{W.~Higemoto$^{1,2,4}$} 
%\author{T.~Matsuzaki$^{5}$}
%\author{Y.~Kobayashi$^6$}
%\author{T.~Sakaguchi$^6$}
%\author{H.~Kageyama$^{6,7,8}$}
%\affiliation{$^1$Advanced Science Research Center, Japan Atomic Energy
% Agency, Tokai, Ibaraki 319-1195, Japan}
%\affiliation{$^2$Materials and Life Science Division, J-PARC Center,
%Tokai, Ibaraki 319-1195, Japan}
%\affiliation{$^3$Institute of Materials Structure Science, High Energy
%Accelerator Research Organization (KEK), Tsukuba,
%Ibaraki 305-0801, Japan}
%\affiliation{$^4$Department of Physics, Tokyo Institute of Technology, 
%Meguro, Tokyo 152-8551, Japan}
%\affiliation{$^5$Advanced Meson Science Laboratory, RIKEN, Wako
%351-0198, Japan}
%\affiliation{$^6$Department of Energy and Hydrocarbon Chemistry,
%Graduate School of Engineering, Kyoto University, Nishikyo-ku,Kyoto
%615-8510, Japan}
%\affiliation{$^7$Department of Chemistry,
%Graduate School of Science, Kyoto University, Sakyo-ku,Kyoto
%606-8520, Japan}
%\affiliation{$^8$Institute for Integrated Cell-Material Sciences, Kyoto
%University, Sakyo-ku, Kyoto 606-8501, Japan}

%\date{\today}

\begin{abstract}
Excited configurations of hydrogen in the oxyhydride BaTiO$_{3-x}$H$_x$ 
 ($x=0.1-0.5$), which are considered to be involved in its hydrogen
 transport and exchange processes, were investigated by positive muon spin
 relaxation ($\mu^+$SR) spectroscopy using muonium (Mu) as a 
 pseudoisotope of hydrogen. 
 Muons implanted into the BaTiO$_{3-x}$H$_x$ perovskite lattice were
 mainly found in two qualitatively different metastable states. One was
 assigned to a highly mobile interstitial protonic state, which is commonly
 observed in perovskite oxides.
 The other was found to form an entangled two spin-$\frac{1}{2}$ system
 with the nuclear spin of an H$^-$ ion at the anion site.
%at the distance of 1.64(1)~\AA. 
The structure of the (H,Mu) complex agrees well with that of a neutralized
  center containing two H$^-$ ions at a doubly charged oxygen vacancy,
 which was predicted to form in the SrTiO$_{3-\delta}$ perovskite lattice by a
 computational study  [Y. Iwazaki {\it et al.}, APL Materials {\bf 2},
 012103 (2014)]. Above 100 K, interstitial Mu$^+$ diffusion and
 retrapping to a deep defect were observed, which could be a rate-limiting
 step of macroscopic Mu/H transport in the BaTiO$_{3-x}$H$_x$ lattice.

% coexistence of two hydrogen species H$^+$/H$^-$ is possibly allowed in
% this system. 

\end{abstract}


\maketitle

%%%%%%%%%%%%%%%%%%%%%%%%%%%%%%%%%%%%
%%%%%%%%%%%%INTRODUCTION%%%%%%%%%%%%
%%%%%%%%%%%%%%%%%%%%%%%%%%%%%%%%%%%%
%\section{Introduction}

%have stimulated a renewed interest in 
%under debate.

Oxyhydrides of perovskite titanates $A$TiO$_{3-x}$H$_x$ ($A$:
Ba, Sr, Ca) have attracted much attention because of their fascinating
properties, which are associated with the lability of H$^-$
ions~\cite{kobayashi12,yajima12,masuda05}. 
The perovskite oxyhydrides are obtained as O$^{2-}$/H$^-$ solid solutions 
from $A$TiO$_3$ by CaH$_2$ reduction.
A combined analysis of X-ray and neutron diffraction data on
BaTiO$_{3-x}$H$_x$ ($x<0.6$) indicates that O$^{2-}$ is randomly
substituted by H$^-$ without creating any detectable amount of vacancies in
the anion sublattice.
This is in sharp contrast with the established hydrogen
configuration in the $A$TiO$_3$ lattice, namely, interstitial protonic
hydrogen H$_i^+$ bound to an O$^{2-}$ ion.
The O$^{2-}$/H$^-$ substitution in BaTiO$_3$ suppresses its structural
transitions and ferroelectricity; a cubic symmetry is maintained at
least down to 2 K.
Macroscopic gas analysis revealed that the hydrogen in $A$TiO$_{3-x}$H$_x$
is mobile and exchangeable in its gaseous environment at a relatively
moderate temperature of $\sim$400$^{\circ}\mathrm{C}$. Furthermore, 
CaH$_2$ reduction changes the parent band insulators into paramagnetic
metals as evidenced by electrical resistivity and magnetic
susceptibility measurements. 
The transport and hydrogen exchange characters of these materials make
them potentially suitable for application in mixed electron/hydrogen
ion conductors and hydrogen membranes.

Several theoretical works have been published to date on the stability of
hydrogen ions in oxygen-deficient perovskite titanates
$A$TiO$_{3-\delta}$ and their transport and exchange
mechanisms~\cite{iwazaki10,iwazaki14,zhang14,varley14}. These studies
commonly concluded that the most stable single hydrogen configuration in
an n-type carrier-rich environment is H$^-$ at a doubly charged oxygen
vacancy V$_{\rm O}^{2+}$, formally expressed as H$_{\rm O}^+$. 
On hydrogen kinetics starting from the H$_{\rm O}^+$ configuration,
two types of scenarios were proposed. One is based on the idea of
correlated migration of H$_{\rm O}^+$, V$_{\rm O}^{2+}$, and 
O$_{\rm O}^{0}$ in the network of the anion site~\cite{kobayashi12}. The other
involves charge-state transitions between H$_{\rm O}^+$ (hydridic) and H$^+_i$ (protonic).
Releasing two electrons to the conduction band, the hydride in
the H$_{\rm O}^+$ center can be transformed to the highly mobile
H$_i^+$, and it then diffuses as a proton from one interstitial site to
another separated by a low potential
barrier~\cite{iwazaki10,zhang14}. After that, it finds another V$_{\rm
O}^{2+}$ center and converts back to H$_{\rm O}^+$ with two electrons.
This scenario explains the hydrogen transport and exchange abilities of 
$A$TiO$_{3-x}$H$_x$ well, without involving migration of the other heavy
elements.
Some calculations dealt with an interaction between an H$_{\rm_O}^+$
center and an incoming H$_i^+$, which should be taken into account when
considering hydrogen transport and exchange in highly hydrogenated
$A$TiO$_{3-x}$H$_x$~\cite{iwazaki14,zhang14}. Three metastable
configurations were theoretically proposed for two hydrogen atoms at the
V$_{\rm O}^{2+}$ center. The simplest one is an H$_2$ molecule-like
state, labeled as (H$_2$)$_{\rm O}^{2+}$. The others are
antisymmetric (2H)$_{\rm O}^0$ and symmetric (2H)$_{\rm O}^{\prime 0}$
states involving two hydrogen ions in the hydride form, proposed by
Iwazaki {\it et al.} in SrTiO$_{3-\delta}$\cite{iwazaki14}.
%These neutralized centers involve two hydrogen ions in the hydride
%form and could be energetically more favored then (H$_2$)$_{\rm
%O}^{2+}$ in a carrier-rich environment. 
These centers could be converted to and back from an (H$_i^+$ + H$_{\rm
O}^+$) defect combination with relatively low activation energy. Because
the H$_i^+$ center is highly mobile, one can consider another path for
hydrogen transport and exchange in $A$TiO$_{3-x}$H$_x$ via the metastable
(H$_2$)$_{\rm O}^{2+}$, (2H)$_{\rm O}^0$, or (2H)$_{\rm O}^{\prime 0}$
states.

%Furthermore, BaTiO$_{3-x}$H$_x$ opens new multistep low temperature
%topochemical routes to access various mixed anion compounds, where H$^-$ works
%as a labile ligand in solid state chemistry\cite{masuda05}. 

In contrast to theoretical advancements, experimental insights into 
the mechanisms of hydrogen transport and exchange in $A$TiO$_{3-x}$H$_x$
are still quite limited.
In this paper, we report on an experimental investigation of excited
hydrogen configurations in the BaTiO$_{3-x}$H$_x$ lattice by the
positive muon spin relaxation ($\mu^+$SR) method, which would be
associated with the hydrogen kinetics in it.
The $\mu^+$SR technique has been widely utilized for the study of local
electronic structures of hydrogen impurities in condensed matter.
Spin-polarized positive muons implanted into a solid 
lose their kinetic energy mainly by ionization and are then trapped at
local potential minima, not necessarily at the global minimum.
The electronic structure of muonium (a hydrogen-like $\mu^+$-$e^-$ bound
state: Mu) is supposed to be very similar to that of
hydrogen except for a small correction due to the difference in reduced
mass ($\sim 0.4$\%). 
The as-implanted mixture of Mu states is far from equilibrium and can 
involve metastable excited states~\cite{lichti08}.
In the BaTiO$_{3-x}$H$_x$ lattice, most implanted muons are
expected to mimic excited configurations of hydrogen together with H$^-$
in the host lattice through their lifetime.
This is because the cross section to produce a Mu analog of the most
stable H$_{\rm O}^+$ center should be quite small owing to the low concentration
of vacancies in the anion sublattice.
%This is because there are not enough vacancies in the anion sublattice
%below 400$^{\circ}\mathrm{C}$ to form a Mu analog of the
%most stable H$_{\rm O}^+$ center. 
%before their decay with the mean lifetime of $\sim$2.2~$\mu$s.
Information on local electronic structures and atomic
configurations around the muons can be obtained from the
time evolution of muon spin polarization, which is driven by 
magnetic interactions between a muon spin and surrounding nuclear and
electron spins.

%%%%%%%%%%%%%%%%%%%%%%%%%%%%%%%%%%%%
%%%%%%%%%%%%EXPERIMENTAL%%%%%%%%%%%%
%%%%%%%%%%%%%%%%%%%%%%%%%%%%%%%%%%%%
%\section{Experimental Methods}
Powder samples of BaTiO$_{3-x}$H$_x$ ($x$=0.1, 0.2, 0.3, and 0.5$\pm
0.05$) with an average grain diameter of 0.17~$\mu$m were prepared from
BaTiO$_3$ powders by a CaH$_2$ reduction method~\cite{kobayashi12}.
$\mu^+$SR experiments were carried out using the D$\Omega$1
spectrometer in the D1 area, J-PARC MUSE, Japan, and the Argus
spectrometer in the Port 2, RIKEN-RAL, UK.
The BaTiO$_{3-x}$H$_x$ powder was compressed into a pellet
% and wrapped with aluminum foil with a thickness of 12~$\mu$m. The
% pellet
and glued with Apiezon N grease on a silver sample holder that was
mounted on a conventional $^4$He-flow cryostat. 
Pulsed $\mu^+$SR measurements were performed in a zero applied field
(ZF) and longitudinal magnetic fields (LF) over the temperature range 15-450~K.
 A double-pulsed surface muon beam was incident to the sample. The
time interval between two muon bunches in each spill was 600~ns for
J-PARC MUSE and 324~ns for RIKEN-RAL. 
The time evolution of muon spin polarization $P(t)$ was extracted from
the forward-backward asymmetry of positrons emitted during the decay of
positive muons. The origin of the time $t$ was set to the arrival time
of the second bunch and data after $t=0$ were analyzed. The influence of the
double-pulsed time structure on $P(t)$ was taken into account according
to the treatment given in Ref.~\cite{Ito10}.

\begin{figure}
\includegraphics[scale =0.65]{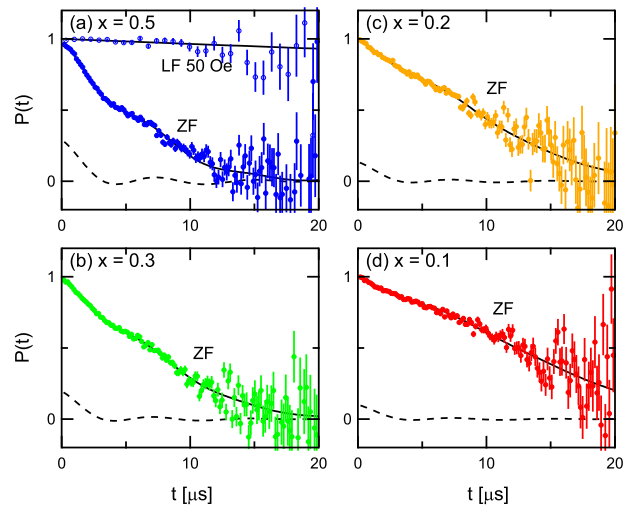}
  \caption{\label{fig1} ZF-$\mu^+$SR spectra of (a) $x=0.5$, (b)
  $x=0.3$, (c) $x=0.2$, and (d) $x=0.1$ samples at 15~K. 
  The solid curves show the best fits to Eq.~(\ref{eq_asym}). Partial
  contribution from the 2S component is displayed by the broken curves.
  A LF-$\mu^+$SR spectrum of the $x=0.5$ sample in a LF of 50~Oe was
  also shown in Fig.~\ref{fig1}(a) together with an exponential fitting
  curve.}
 \end{figure}

%%%%%%%%%%%%%%%%%%%%%%%%%%%%%%%%%%%%
%%%%%%%%%%%%%%RESULTS%%%%%%%%%%%%%%%
%%%%%%%%%%%%%%%%%%%%%%%%%%%%%%%%%%%%
%\section{Results and Discussion}
%\subsection{$x$ dependence at 15~K}
We first discuss the $x$ dependence of ZF-$\mu^+$SR spectra for  
BaTiO$_{3-x}$H$_{x}$. Figure~\ref{fig1} shows a ZF-$\mu^+$SR
spectrum of as-prepared BaTiO$_{3-x}$H$_{x}$ samples at 15~K, together
with an LF-$\mu^+$SR spectrum of the $x=0.5$ sample at 50~Oe.
The constant background from muons that missed the sample was identified by
using a reference sample and has been subtracted from the spectra. 
There was no sign of paramagnetic Mu$^0$ in the ZF-$\mu^+$SR spectra of
BaTiO$_{3-x}$H$_{x}$ in contrast to the parent band insulator BaTiO$_3$, where
shallow-donor Mu$^0$ was detected below $\sim$80~K~\cite{ito13}. 
The absence of Mu$^0$ in BaTiO$_{3-x}$H$_{x}$ is
consistent with its metallic conductivity, since conduction
electrons screen the positive charge of $\mu^+$ and prevent an electron from
localizing around it.

Muon spin relaxation in ZF is mainly caused by quasi-static
magnetic interactions between a diamagnetic muon (Mu$^+$ or Mu$^-$) and
surrounding nuclei, since the polarization is mostly recovered by
applying a small LF of 50~Oe, as shown in Fig.~\ref{fig1}(a).
The muon spin depolarization curves in ZF have an oscillating feature
superposed on a Gaussian relaxation curve. Such spontaneous oscillation 
is usually regarded as evidence of coherent magnetic
order. However, this is clearly not the case, since there is no sign 
of magnetic order in the temperature dependence of magnetic
susceptibility down to 2~K~\cite{kobayashi12}.
Strong coupling between a muon spin and a small
number of nearby nuclear spins also causes such an oscillation in
ZF-$\mu^+$SR spectra even in nonmagnetic or paramagnetic
materials~\cite{brewer86,nishiyama03,lancaster07,lord00,kadono08,sugiyama10,ito09}.
In BaTiO$_{3-x}$H$_{x}$, the oscillating component is ascribed to the
coupling between a muon spin ($S=\frac{1}{2}$) and a nearby $^1$H nuclear spin
($I=\frac{1}{2}$), since $^1$H has a large nuclear dipole 
moment in comparison with those of stable Ba and Ti nuclei~\cite{stone05}.
Considering the low concentration of $^1$H, we ignore the situation where 
more than two $^1$H nuclear spins equally couple to a muon spin as a 
collinear $^{19}$F-$\mu^+$-$^{19}$F spin
configuration~\cite{brewer86,nishiyama03}.
The muon spin relaxation function for the entangled two
spin-$\frac{1}{2}$ state (hereafter, referred to as 2S) is expressed as
\begin{align}
G_{\rm 2S}(t)&=\frac{1}{6}+\frac{1}{6}\cos(2\pi f_d t)\nonumber\\
&+\frac{1}{3}\cos(\pi f_d t)+\frac{1}{3}\cos(3\pi f_d t),\label{eq_2G}\\
&f_d=\frac{\mu_0\hbar\gamma_{\mu}\gamma_{I}}{8\pi^2 d^3},\label{eq_fd}
\end{align}
where $d$ is the distance between $\mu^+$ and $^1$H at the anion site
and $\gamma_{\mu}$ and $\gamma_{I}$ are the gyromagnetic ratios of $\mu^+$
and $^1$H ($\gamma_{\mu}/2\pi=135.53~{\rm MHz/T}$ and
$\gamma_{I}/2\pi=42.58~{\rm MHz/T}$),
respectively~\cite{nishiyama03,lancaster07}.
%The 2S state muon must locate considerably near an H$^-$ ion
%to make the dipolar interaction with the nearby $^1$H nucleus
%sufficiently stronger than those with further nuclei.
On the other hand, the Gaussian relaxation component is ascribed to 
diamagnetic muons that do not create such a special magnetic bond with a single
$^1$H. These muons should be located further from H$_{\rm O}^+$ centers
than those in the 2S state and lose their spin polarization via
magnetic dipolar interactions with a large number of surrounding nuclei.
In such a case, the Gaussian Kubo-Toyabe relaxation function
(relaxation rate $\Delta$) based on local field approximation is
useful~\cite{hayano79} and its early-time region ($t<2\Delta^{-1}$) is
well approximated by the Gaussian function $e^{-\Delta^2 t^2}$.

\begin{figure}
\includegraphics[scale =0.59]{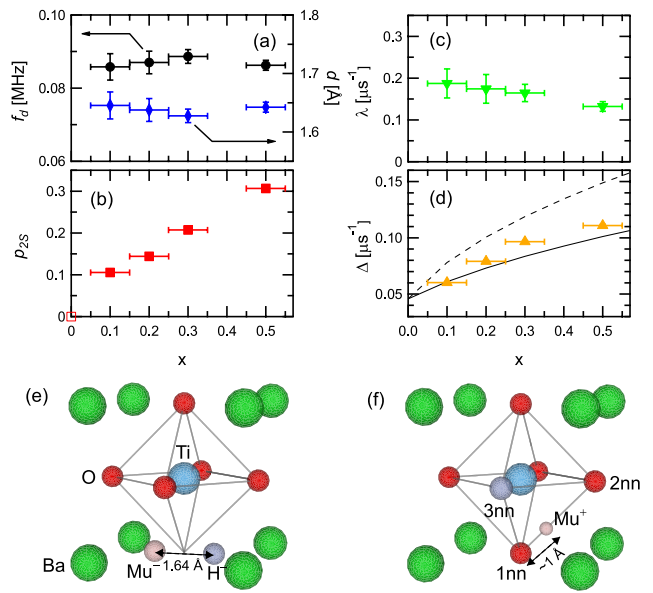}
  \caption{\label{fig2} $x$ dependences of (a) $f_d$ and $d$, (b) $p_{\rm
  2S}$, (c) $\lambda$, and (d) $\Delta$ at 15~K in ZF, and the most probable
  atomic configurations associated with (e) the 2S component and (f) the
  Gaussian component.}
 \end{figure}

The ZF-$\mu^+$SR spectra at 15~K were fitted to the following function, 
including the 2S and Gaussian components, i.e., 
\begin{equation}
P(t)=p_{\rm 2S}e^{-\lambda t}G_{\rm 2S}(t) + (1-p_{\rm
 2S})e^{-\Delta^2t^2}, \label{eq_asym}
\end{equation}
where $\lambda$ and $p_{\rm 2S}$ are the relaxation rate and the
fraction of the 2S state, respectively. 
The exponential function phenomenologically describes spin relaxation of 
the 2S state~\cite{sugiyama10}, which can be caused by inhomogeneity,
surrounding nuclei, and slow dynamics.
Satisfactory fits were obtained, as shown in Fig.~\ref{fig1} with solid
curves. Figure~\ref{fig2} shows the $x$ dependences of $f_d$, 
$p_{\rm 2S}$, $\lambda$, and $\Delta$, obtained from the fits. The
distance of 2S spins, $d$, is also shown in Fig.~\ref{fig2}(a),
calculated from $f_d$ via Eq.~(\ref{eq_fd}). 
The linear increase in $p_{\rm 2S}$ suggests that the formation probability
of the 2S state is proportional to the hydrogen concentration.
The constant $d$ is reasonable because the local structure of the 2S state
should be independent of $x$ for such low hydrogen concentrations. 
The average length, 1.64(1)\AA, is much shorter than the distance 
between two nearest-neighbor (nn) anion sites, $\sim$2.8~\AA~and much
longer than the bond length of a hydrogen molecule, 0.74~\AA. This indicates that
the 2S state can be assigned neither to a (Mu$_{\rm O}^+$, H$_{\rm O}^+$)
defect combination over two nn anion sites, nor a molecular MuH center at
a doubly charged oxygen vacancy. On the other hand, the average value of
$d$ agrees well with the theoretical H-H distances for the
(2H)$_{\rm O}^{\prime 0}$ and (2H)$_{\rm O}^0$ centers in
SrTiO$_{3-\delta}$, 1.67 and 1.64~\AA, respectively~\cite{iwazaki14}. Because
the Ba and Sr compounds have almost the same lattice constant, it seems 
reasonable to ascribe the 2S state to one of the Mu$^-$ analogs of these
centers, namely, (Mu,H)$_{\rm O}^{\prime 0}$ or (Mu,H)$_{\rm O}^0$, as
shown in Fig.~\ref{fig2}(e).
Our observation demonstrates that the uncommon (2H)$_{\rm
O}^{\prime 0}$ or (2H)$_{\rm O}^0$ center, where the hydrogen exchange can
occur, is really allowed to form in the n-type perovskite lattice at
least as a metastable state.
This center is also important in carrier-control technology for transition-metal
oxides because the V$_{\rm O}^{2+}$ donor can be completely passivated by
forming this with hydrogen~\cite{iwazaki14}.

%where both H and Mu are anionic.
%It also seems difficult to associate this with an interstitial Mu$_i^+$,
%which is supposed to form a covalent bond with an O$^{2-}$ ion in the
%perovskite lattice. Assuming that the interstitial Mu$_i^+$ locates at
%the same position of the H$_i^+$ center\cite{iwazaki10,zhang14}, the
%distance to the nearest O ion other than the covalent O ion is 1.8
%The electrostatic repulsion between the Mu$_i^+$ and the H$_{\rm
%O}^+$ centers might make their distance even longer.

%Since the averaged $d$ is much longer than the bond length of an H$_2$
%molecule, one can conclude that the muon implanted in BaTiO$_{3-x}$H$_{x}$
%preferably forms an O-$\mu$ bond rather than an H-$\mu$ bond with a
%substitutional H$^-$.
The $\Delta$ included in the Gaussian component monotonically increases with
increasing $^1$H concentration, as shown in Fig.~\ref{fig2}(d). This
behavior can be quantitatively explained by assigning the Gaussian
component to a Mu$^+$ analog of the H$^+_i$ center
bound to an O$^{2-}$ ion. Here we assume a simplified atomic
configuration as illustrated in Fig.~\ref{fig2}(f); Mu$^+_i$ is placed on
an edge of the anion octahedron in the undistorted perovskite
lattice with a lattice constant of 4.02~\AA~\cite{kobayashi12} and 
the O-Mu bond length is set to 0.987~\AA, based on a first-principles
calculation on the structure of the H$^+_i$ center in cubic
BaTiO$_3$~\cite{iwazaki10}. The first nn anion is fixed to be O$^{2-}$ to form
the O-Mu bond and O$^{2-}$ at the third nn and further anion sites is
randomly replaced by H$^-$ with a probability of $x/3$. 
Under these conditions, we calculated the rms width
$\sigma(=\Delta/\gamma_{\mu})$ of the local field distribution at the
Mu$^+_i$ site for the following two cases with regard to the second nn anion
configuration: (i) always O$^{2-}$, (ii) randomly replaced by H$^-$
with a probability of $x/3$.
The $\Delta(x)$ curves for cases (i) and (ii) were obtained from the
relation: $\Delta=\gamma_{\mu}\sigma$, as shown in Fig.~\ref{fig2}(d)
with solid and broken lines, respectively. While both curves reproduce
the increasing trend of the experimental data, the agreement is even
better for case (i). This suggests that a nearly collinear
O-Mu$\cdots$H configuration (H$^-$ at the 2nd nn anion site) is unstable 
and, even if it forms, it immediately falls into the (Mu,H)$_{\rm
O}^{\prime 0}$ or (Mu,H)$_{\rm O}^0$ state, as shown in
Fig.~\ref{fig2}(e). 

%while the first nn anion site is always filled with
%O$^{2-}$ to form the O-Mu bond.

%\subsection{Assignment of muon localization sites at 15~K}

\begin{figure}
 \includegraphics[scale =0.7]{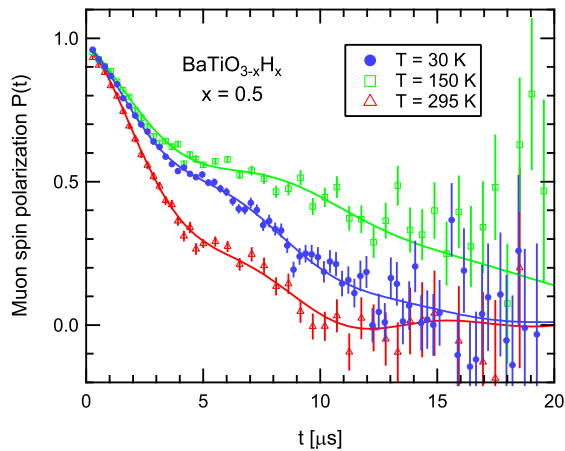}
  \caption{\label{fig3} ZF-$\mu^+$SR spectra of BaTiO$_{2.5}$H$_{0.5}$
  ($x=0.5$) at 30, 150, and 295~K. 
  The solid curves show the best fits to Eq.~(\ref{eq_asym}).}
 \end{figure}

\begin{figure}
 \includegraphics[scale =0.59]{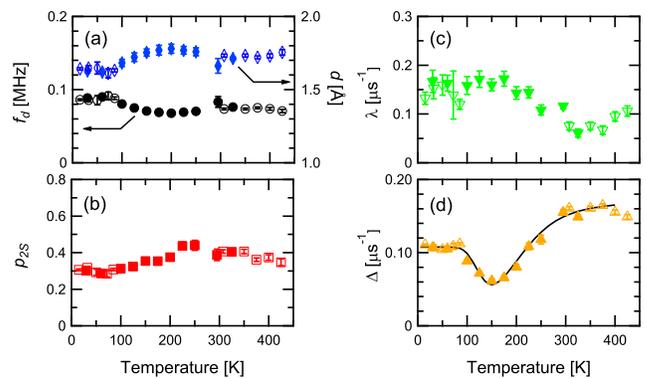}
  \caption{\label{fig4} Temperature dependences of (a) $f_d$ and $d$, (b) $p_{\rm
  2S}$, (c) $\lambda$, and (d) $\Delta$ in ZF for
  BaTiO$_{2.5}$H$_{0.5}$. The solid curve in (d) is the best fit to the
  simplified two-state model (see text).}
 \end{figure}

We then move on to the temperature dependence of ZF-$\mu^+$SR
spectra in BaTiO$_{2.5}$H$_{0.5}$ ($x=0.5$). Figure~\ref{fig3} shows
ZF-$\mu^+$SR spectra of the as-prepared sample at 30, 150, and
295~K. The spectrum rises with increasing temperature from 30 to
150~K and then falls as a result of further warming to 295~K. In
spite of the peculiar temperature
dependence, the two-component feature (2S and Gaussian) seems to be
retained within the temperature range of our experiment. We therefore used
Eq.~(\ref{eq_asym}) to fit the ZF spectra and obtained
satisfactory fits, as shown in Fig.~\ref{fig3} with solid curves. 
The parameters $f_d$, $d$, $p_{\rm 2S}$, $\lambda$, and $\Delta$
obtained by the fits are shown in Fig.~\ref{fig4} as functions of
temperature. Closed and open symbols show data taken before and after
keeping the sample at 450~K for four hours in a cryostat vacuum,
respectively. 
No significant influence of the four-hour annealing is observed in 
Fig.~\ref{fig4}, indicating that the H$_{\rm O}^+$ configuration in the
host lattice was maintained during the experiment and thus the temperature
variation of the spectra should be primarily attributed to muon
kinetics.

Moderate temperature dependences in $d$ and $p_{\rm 2S}$ for the 2S
component indicates that conversion between the 2S and the Mu$_i^+$
states seldom occurs below 450~K. This suggests that higher temperature
is necessary to activate the hydrogen exchange process proposed by Zhang
{\it et al.}~\cite{zhang14} 
On the other hand, the $\Delta$ for the Gaussian component shows a notable
temperature dependence with double
plateaus separated by a concave at around 150~K. Similar behavior was 
reported in Sc-doped SrZrO$_3$, explained within the
framework of a two-state model on interstitial Mu$^+$
diffusion~\cite{hempelmann98,borghini78}. Here, we adopted a simplified
version of this model, as was used in Ref.~\cite{lord01}.
In this framework, muons are assumed to start in the Mu$_i^+$ site 
with a static Gaussian width $\Delta_0$ and diffuse between equivalent
sites with activation energy $E_a$ and attempt rate $\nu_0$. They are
then captured with a probability $c$ par hop at a deep trap site
with a static Gaussian width $\Delta_1$ and stay there until they decay. 
We numerically evaluated $\Delta(T)$ as a function of $\Delta_0$,
$\Delta_1$, $E_a$, $\nu_0$, and $c$, and fitted it to
the experimental $\Delta$ in the temperature range 15-400~K with $\Delta_0$
fixed at 0.108 $\mu$s$^{-1}$ (the average value of $\Delta$ below
85~K). Consequently, we obtained the best fit curve, as shown in
Fig.~\ref{fig4}(d), and the following parameters: 
$\Delta_1=0.175(3)~\mu$s$^{-1}$, $E_a=0.086(4)$~eV, $\nu_0=10(3)\times
10^8$~s$^{-1}$, and $c=0.025(4)$. The value of $E_a$ is close 
to that in Sc-doped SrZrO$_3$~\cite{hempelmann98}, demonstrating the
validity of our analysis. This is about half the calculated potential
barrier height of $\sim 0.2$~eV for interstitial diffusion of protons in
cubic BaTiO$_3$~\cite{zhang14}. The difference could be the consequence of
a higher vibrational zero-point energy for the
muon~\cite{norby90,hempelmann98}. The increase in $\Delta$ above 150~K
corresponds to the trapping of the highly mobile Mu$^+_i$ at
the deep trap site with a static Gaussian width $\Delta_1$, which is 
considerably larger than $\Delta_0$. This might be a rate-limiting step of
macroscopic Mu/H conduction in BaTiO$_{3-x}$H$_{x}$. The large
$\Delta_1$ suggests a high density of hydrogen around the deep trap and
thus a Mu capture at V$_{\rm O}^{2+}$ is unlikely.
%but this has not yet been confirmed. 

It should be noted that our methodology cannot deal with the case of direct
H$^-$ migration in the network of anionic
octahedra~\cite{kobayashi12}. Our experimental results
never exclude such a process, which is considered to be more or less
activated at high temperature and contribute to net H transport together
with other possible processes.

In conclusion, we investigated excited hydrogen configurations allowed in
the BaTiO$_{3-x}$H$_{x}$ lattice by the $\mu^+$SR spectroscopy.
% using Mu as a pseudoisotope of H. 
Our experimental results are mostly in line with a theoretical proposal
on hydrogen exchange and transport involving charge-state
transitions~\cite{iwazaki10, zhang14,iwazaki14},
but also indicate the existence of a deep Mu/H trap and its importance
in macroscopic Mu/H transport. It remains for future studies to fully
identify this deep trap.

\begin{acknowledgments}
We thank T.~Sakaguchi and D.~Tomono for technical assistance and Y.~Iwazaki and
 K.~Nishiyama for helpful discussions. This work was supported by CREST,
 JST and the MEXT KAKENHI Grant No. 24108509 and the JSPS KAKENHI Grant
 No.~16K17544.
\end{acknowledgments}

\end{document}